\documentstyle[amssymb,aps,epsfig,prl,twocolumn]{revtex}

\begin{document}
\title{{\LARGE Electronic transport through nuclear-spin-polarization-induced
quantum wire}}
\author{\-{\it Yu.V. Pershin}$^{1,4}${\it , S.N. Shevchenko}$^{1,2}${\it , I.D. Vagner}%
$^{1,3,4}${\it \ and P.Wyder}$^{1}$}
\address{$^{1}${\small Grenoble High Magnetic Fields Laboratory, \ \ }\\
{\small Max-Planck-Institute f\"{u}r Festkorperforschung and CNRS,\ \ }\\
{\small BP 166, F-38042 Grenoble Cedex 9, France}\\
{\small \ }$^{2}${\small B.I.Verkin Institute for Low Temperature Physics}\\
{\small and Engineering,\ }\\
{\small 47 Lenin Avenue 61103 Kharkov, Ukraine}\\
{\small \ }$^{3}${\small Holon Academic Institute of Technology, \ }\\
{\small 52 Golomb St., Holon 58102 Israel}\\
$^{4}${\small Department of Physics, Clarkson University,}\\
{\small Potsdam, NY 13699-5820, USA.}}
\maketitle

\begin{abstract}
Electron transport in a new low-dimensional structure - the nuclear spin
polarization induced quantum wire (NSPI QW) is theoretically studied. In the
proposed system the local nuclear spin polarization creates the effective
hyperfine field which confines the electrons with the spins opposite to the
hyperfine field to the regions of maximal nuclear spin polarization. The
influence of the nuclear spin relaxation and diffusion on the electron
energy spectrum and on the conductance of the quantum wire is calculated and
the experimental feasibility is discussed.
\end{abstract}


There has been much recent theoretical \cite{VRWZ98,CSRV99,FIPV01} and
experimental \cite{Berg,Kane,Wald,Dietsche} interest in the peculiarities of
the electron transport in mesoscopic systems with highly polarized nuclear
spins. In \cite{VRWZ98} a new class of phenomena: the {\bf %
meso-nucleo-spinics} was proposed based on the strong influence of
the hyperfine magnetic field of the polarized nuclei on the energy
spectrum and the wave functions of the conduction electrons in few
channel mesoscopic systems. It was shown there and later discussed
in \cite{CSRV99} that the inhomogeneous external hyperfine
magnetic field, acting on the charge carriers confined to move in
a ring, influence the quantum interference (mesoscopic) phenomena
and can induce the persistent current with some interesting
physical features. In \cite{FIPV01} the electron states in a
quantum dot with a nuclear spin polarization were studied within
the perturbation theory. Moreover, it was proposed to use the
inhomogeneous magnetic field to create so-called magnetic
structures, such as magnetic quantum dots, rings, superlattices
etc. (see for a review Ref. \cite{Kim}).

The conductance quantization, in the case of the ballistic transport through
quantum wires at low temperatures, is by now well studied both
experimentally and theoretically \cite{Wees88,Wharam88,Datta}. The
dependence of the conductance at 'zero' temperature on the number of
transverse modes in the conductor is given by the Landauer formula \cite
{Landauer85}:

\begin{equation}
G=\frac{2e^{2}}{h}MT\text{ \ , }  \label{G}
\end{equation}
where $T$ is the average electron transmission probability, $M$ is the
number of the transverse modes and the factor $^{\prime }2^{\prime }$ stands
because of the spin degeneration. It is assumed that the transition
probability $T$ is independent of the energy in a small interval between the
chemical potentials of the reservoirs. Usually, the number of the transverse
modes, defined by the effective width of the conductor, is controlled by the
gate voltage and the conductance is changed in discrete steps $\frac{2e^{2}}{%
h}$ \cite{Datta}.

In this paper we propose a new class of experiments, based on the creation
of a quantum wire by the hyperfine magnetic field of polarized nuclei acting
on the conduction electron spins. The time evolution of the hyperfine
magnetic field, due to the nuclear spin relaxation and the nuclear spin
diffusion, leads to variation of the number of transverse modes and
corresponding electron energies at a constant gate potential that can be
directly measured by transport experiments.

The proposed system is depicted on Fig.1. The two dimensional
electron gas (2DEG) is splitted in two parts by a potential
applied to the gate electrode (of the width $L$) located under
2DEG. Two parts of 2DEG are connected by a NSPI QW created either
by the optical nuclear spin polarization \cite
{Lampel,Zahar,Barrett}, or by the transport polarization
\cite{Kane,Wald} or by any other suitable experimentally method.
The chemical potentials of these 2DEG regions are $\mu _{1}$ and
$\mu _{2}$, $\frac{\mu _{1}-\mu _{2}}{\left| e\right| }$ is the
bias voltage. We assume that the bias voltage is small in
comparison with the potential which confines the electrons. The
conductance of the NSPI QW depends on the number of the transverse
modes below $\mu _{1}$ and $\mu _{2}$. Since the number of the
transverse modes changes in time, the current between two regions
of 2DEG \ is also time-dependent.

The contact hyperfine interaction between an electron and nuclear
spins is described by the Fermi contact Hamiltonian
\cite{Slichter} $H_{hf}\left(
{\bf r}\right) =\frac{8\pi }{3}\mu _{B}\gamma _{n}\hbar \sum_{i}{\bf I}_{i}%
{\bf \sigma }\delta \left( {\bf r}-{\bf r}_{i}\right) $,
\begin{figure}[t]
 \centering
 \includegraphics[width=8cm]{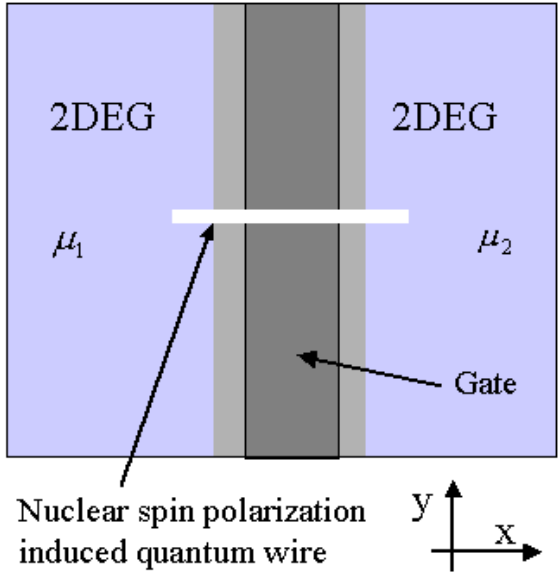}
 \caption{The geometry of the proposed experiment: the 2DEG is splitted by the
gate potential, the narrow conductor is created by the nuclear
spin polarization.}
\end{figure}
where $\mu _{B}$ is the Bohr magneton, ${\bf I}_{i}$ and ${\bf
\sigma }$ are nuclear and
electron spins, ${\bf r}_{i}$ and ${\bf r}$ are the radius vectors of the $i$%
-th nucleus and of the electron. It follows that once the nuclear spins are
polarized, the charge carrier spins feel the effective hyperfine field ${\bf %
B}_{hf}$\ which lifts the spin degeneracy. In $GaAs/AlGaAs$-heterojunction
one may achieve the hyperfine field of several Teslas \cite{Wald,Berg}. The
spin splitting ($\mu _{B}B_{hf}$) due to such hyperfine magnetic field is
comparable to the Fermi energy of 2DEG. Thus, if the gate potential $%
U_{gate} $ is more than $\mu _{1,2}$, then all the electrons in the region,
where nuclear spins are polarized, will occupy the energetically more
favorable states with the spins opposite to ${\bf B}_{hf}$. And,
furthermore, the nuclear polarization acts on the electrons as the effective
confining potential $V_{conf}=-\mu _{B}B_{hf}$. This effective confining
potential can be used to create different nanostructures with polarized
electrons in them. Here we restrict our consideration to NSPI QW.

There are two main mechanisms leading to the\ time dependence of
the hyperfine field: the nuclear spin relaxation and the nuclear
spin diffusion. We assume that the nuclear spin polarization is
homogeneous in $x$ and $z$ directions. Then the hyperfine field
evolution is described by the one-dimensional diffusion equation
with taking into account the relaxation processes:

\begin{equation}
\frac{\partial B_{hf}}{\partial t}=D\frac{\partial ^{2}B_{hf}}{\partial y^{2}%
}-\frac{1}{T_{1}}B_{hf}\text{,}  \label{diffusion}
\end{equation}
where $D$ is the spin-diffusion coefficient and $T_{1}$ is the
nuclear spin relaxation time \cite{Wolf}. Here we assume that the
nuclear spin polarization is inhomogeneous across the NSPI QW and
quite homogeneous along it (at least in depleted region), provided
that the length of nuclear spin polarized region is larger then
the depleted region. In this case the diffusion in x direction
from ends is irrelevant for the properties of NSPI QW.

Let us discuss experimental feasibility of this assumption. Using
the method of optical nuclear spin polarization
\cite{Lampel,Zahar,Barrett}, the sample is illuminated locally by,
for example, putting a mask on it. The resolution of the optical
illumination of the sample can be high enough. Usual optic
technique allows to create the light beams of the width of the
order of the wave length ($\sim 500nm$), by using near fields
optics the beam width can be sufficiently reduced ($\sim 100nm$).
Hence a NSPI QW of the width of $1 \mu m$ can be created by the
modern experimental technique. In semiconductor heterostructures
having supreme quality, the electron mean free path can be as
large as $100 \mu m$ and NSPI QW will operate in the quantum
regime. We assume the initial condition to be of the Gaussian
form: $B_{hf}\left( y,0\right) =B_{0}\exp \left(
-\frac{y^{2}}{2d^{2}}\right) $. The two parameters, $d$ and
$B_{0}$, define the half-width and the amplitude of the initial
distribution of the hyperfine field, respectively. Then the
solution of Eq.(\ref{diffusion}) is:

\begin{equation}
B_{hf}\left( y,t\right) =B_{0}e^{-\frac{t}{T_{1}}}\left( 1+\frac{t}{t_{0}}%
\right) ^{-\frac{1}{2}}e^{-\frac{y^{2}}{2d^{2}\left( 1+\frac{t}{t_{0}}%
\right) }}\text{ \ \ ,}  \label{B(r,t)}
\end{equation}
where $t_{0}=\frac{d^{2}}{2D}$.

The microscopic description is based on the following Hamiltonian:

\begin{equation}
H=-\frac{\hbar ^{2}}{2m^{\ast }}\Delta +V\left( x\right) +U\left( z\right)
+\mu _{B}{\bf \sigma B}_{hf}\left( y,t\right)  \label{H}
\end{equation}
Here $m^{\ast }$ is the electron effective mass and $V\left( x\right) $ is
the potential energy associated with the gate. The form of the potential $%
V\left( x\right) $ is not very important in our consideration. It
defines the transmission probability in Eq.(\ref{G}). If we take
$V\left( x\right) =V_{0}=const$ for $\left| x\right| <\frac{L}{2}$
and $V\left( x\right) =0$ otherwise, then $T=1$. We suppose, as it
is usually done for 2DEG, that only the lowest subband,
corresponding to the confinement in $z$ direction, is occupied and
we can ignore the higher subbands. Thus, we omit in the following
$z$-dependence of the wave function. The nuclear-spin relaxation
time $T_{1}$ and the characteristic diffusion time $t_0$ in
semiconductors at sufficiently low temperatures is rather long.
They vary from several hours to few minutes \cite{Zahar}. Thus the
time scale introduced by nuclear spin system is several order of
magnitude larger than the time scale of typical electron
equilibration processes. In such a case the conduction electrons
see a quasi-constant nuclear field. This simplifies calculation by
avoiding the complications which would appear when solving the
Schr\"{o}dinger equation with the time dependence due to polarized
nuclei. Taking into account the electrons only with opposite to
the hyperfine field spins (for which the effective potential is
attractive), we obtain the following equation for the transverse
modes energy spectrum:
\begin{figure}[t]
 \centering
 \includegraphics[width=8cm]{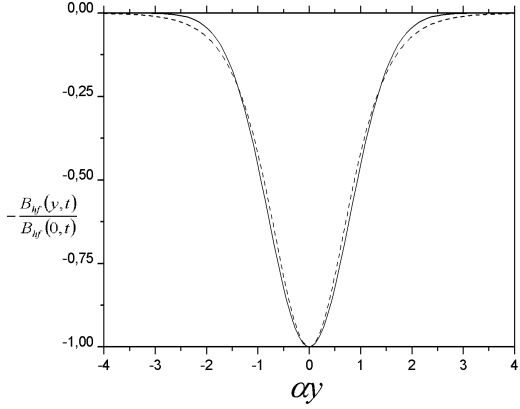}*
 \caption{Comparison of the exact hyperfine field (solid line) and the model
hyperfine field (dashed line).}
\end{figure}
\begin{figure}[t]
 \centering
 \includegraphics[width=8cm]{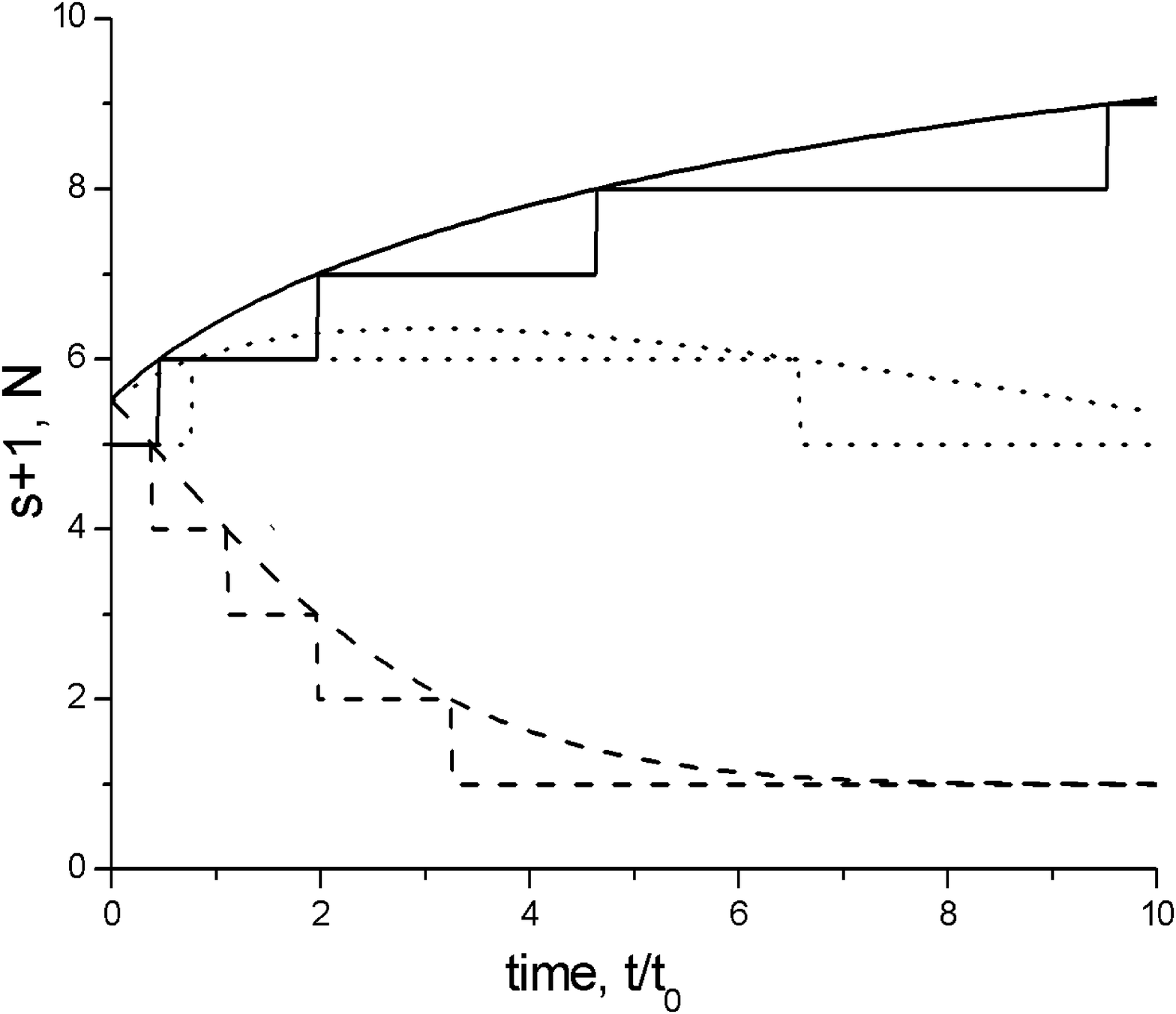}
 \caption{ Different kinds of the level number dependencies on time $N=N\left(
t\right) $ and $s=s\left( t\right) $\ for $C=100$: solid line -
diffusive regime $\left( \frac{T_{1}}{t_{0}}=80\right) $; dashed
line - relaxation regime $\left( \frac{T_{1}}{t_{0}}=1\right) $;
dotted line - intermediate regime $\left(
\frac{T_{1}}{t_{0}}=8\right) $. The stepped lines correspond to
the level number $N$, the smooth lines - $s+1$.}
\end{figure}
\begin{equation}
-\frac{\hbar ^{2}}{2m^{\ast }}\frac{\partial ^{2}\psi \left( y\right) }{%
\partial y^{2}}-\mu _{B}B_{hf}\left( y,t\right) \psi \left( y\right)
=\varepsilon ^{tr}\psi \left( y\right) \text{ .}  \label{transverse eq.}
\end{equation}
To proceed we approximate the hyperfine field (\ref{B(r,t)}) by

\begin{equation}
\widetilde{B}_{hf}=\frac{1}{\mu _{B}}\frac{U_{0}}{\cosh ^{2}(\alpha y)}
\label{new field}
\end{equation}
connected with (\ref{B(r,t)}) by the relations:

\begin{eqnarray}
B_{hf}\left( 0,t\right) &=&\widetilde{B}_{hf}\left( 0,t\right)
\label{connection} \\
\int B_{hf}\left( y,t\right) dy &=&\int \widetilde{B}_{hf}\left( y,t\right)
dy  \nonumber
\end{eqnarray}
It follows from Eq.(\ref{connection}) that both fields at $y=0$ have the
same value, as well as the areas under the curves $B_{hf}\left( y,t\right) $
and $\widetilde{B}_{hf}\left( y,t\right) $ at any fixed $t$. This provides
the total nuclear spin polarization to be the same for the two fields. From
Eqs.(\ref{connection}) we obtain $U_{0}=\mu _{B}B_{0}e^{-\frac{t}{T_{1}}%
}\left( 1+\frac{t}{t_{0}}\right) ^{-\frac{1}{2}}$ and $\alpha ^{-1}=d\sqrt{%
\frac{\pi }{2}}\left( 1+\frac{t}{t_{0}}\right) ^{\frac{1}{2}}$.
From Fig.2 it is clearly seen that the hyperfine fields given by
Eq.(\ref{B(r,t)}) and by Eq.(\ref{new field}) have very similar
dependence on $y$. Substituting the hyperfine field, Eq.(\ref{new
field}), into Eq.(\ref{transverse eq.}), we obtain the
one-dimensional Schr\"{o}dinger equation with the modified
P\"{o}schl-Teller potential:
\begin{equation}
-\frac{\hbar ^{2}}{2m^{\ast }}\frac{\partial ^{2}\psi \left( y\right) }{%
\partial y^{2}}-\frac{U_{0}}{\cosh ^{2}(\alpha y)}\psi \left( y\right)
=\varepsilon ^{tr}\psi \left( y\right) \text{ .}  \label{newShred}
\end{equation}
Solution of Eq.(\ref{newShred}) can be expressed in terms of the
hypergeometric function \cite{Landau} and the energy spectrum is

\begin{equation}
\varepsilon _{n}^{tr}=-\frac{\hbar ^{2}\alpha ^{2}}{2m^{\ast }}\left(
s-n\right) ^{2}\text{ \ \ ,}  \label{transverse energy}
\end{equation}
where $s=\frac{1}{2}\left( -1+\sqrt{1+\frac{8m^{\ast }U_{0}}{\hbar
^{2}\alpha ^{2}}}\right) $ and $n=0,1,2,...$ The number of the energy levels
$N$ is finite and, being defined by the condition $n<s$, is given by $N=1+%
\left[ s\right] $, where $\left[ ...\right] $ denotes the integer
part. A level appears or disappears in the system when $s\left(
t\right) $ becomes an integer.

Let us consider the time dependence of the parameter $s$ in more
detail. The expression for $s$ can be rewritten in a more
convenient form:
\begin{equation}
s\left( t\right) =\frac{1}{2}\left( -1+\sqrt{1+Ce^{-\frac{t}{T_{1}}}\sqrt{1+%
\frac{t}{t_{0}}}}\right)  \label{s}
\end{equation}
where the dimensionless parameter $C=4\pi \frac{m^{\ast }d^{2}}{\hbar ^{2}}%
\mu _{B}B_{0}$ contains only the initial distribution parameters $d$ and $%
B_{0}$. There are two different characteristic times in (\ref{s}): the
diffusion characteristic time $t_{0}$ and the relaxation characteristic time
$T_{1}$. It is easy to see that $s$ has a maximum at $t_{\max }=\frac{T_{1}}{%
2}-t_{0}$. We can distinguish three regimes: the diffusive regime $(T_{1}\gg
t_{0}\sim t)$; the relaxation regime $(T_{1}<2t_{0})$ and the intermediate
regime $(T_{1}\gtrsim 2t_{0})$. Here $t$ is the observation time. Time
dependencies of the number of the energy levels $N$ in different regimes are
shown on Fig.3. In the diffusive regime (full line) the number of levels
increases with time, \ while in the relaxation regime (dashed line) the
number of levels decreases with time. In the intermediate regime (dotted
line) there is the maximum of the function $s\left( t\right) $ at $t=t_{\max
}$. The estimation of the number of the transverse energy levels for the
following set of parameters: \ $B_{0}=1T$, $d=1\mu m$, $m^{\ast }=0.067m_{e}$%
\ \ gives us $N\left( t=0\right) =13$. We can also estimate the number of
electrons per unit length: $n_{L}=\frac{1}{\pi \hbar }\sum \sqrt{2m^{\ast
}\left| \varepsilon _{n}^{tr}\right| }=\frac{\alpha }{\pi }N(s-\frac{1}{2}%
(N-1))$, which at $t=0$ and $N\gg 1$ gives $n_{L}\simeq \frac{1}{\sqrt{2}\pi
^{3/2}}\frac{N^{2}}{d}$. For $N=13$ and $d=1\mu m$ we have $n_{L}\simeq
2\cdot 10^{5}cm^{-1}$.

The time dependence of the transverse energy levels in the diffusive regime
is shown on Fig.4a. The chemical potentials $\mu _{1}$ and $\mu _{2}$ are
depicted by the horizontal lines. The number of transverse energy levels
below them at the moment of time $t$ is the number $M$ of the transverse
modes. The points of intersection of the chemical potentials with the
transverse energy levels are given by

\begin{equation}
\mu _{1,2}=V_{0}-\varepsilon _{n}^{tr}\left( t\right) \text{ \ \ .}
\label{intersaction}
\end{equation}
An interesting feature of the transverse energy spectrum is that the energy
of some levels (for example, the level with $n=3$ in Fig.4a) at the short
times decreases until it has a minimum. In the relaxation regime (Fig.4b)
the absolute values of the energy levels and the number of the levels
decreases monotonically in time. At large times only the energy level with $%
n=0$ survives. In the intermediate regime a mixing of the relaxation and
diffusion regimes happens (Fig.4c): at short times the energy level behavior
is determined by the nuclear spin diffusion and at large times - by the
nuclear spin relaxation.

The dependence of the conductance on the time is shown on Figs.4d-4f. When
an energy level crosses the chemical potentials (the time of the
intersection is given by Eq.(\ref{intersaction})) the conductance changes by
$\frac{e^{2}}{h}$. We underline that due to the spin selective effective
potential the height of the conductance steps is just $\frac{e^{2}}{h}$
which is a half of the conductance quantum $G_{0}=\frac{2e^{2}}{h}$.

It is quite clear that sharp conductance quantization steps shown
on Figs.4d-4f will be smoothed in real experiment. The main
mechanisms of smoothing are: the effect of finite temperature;
scattering of electrons on impurities and defects; and
inhomogeneity of the initial hyperfine field profile along the
wire. The temperature smoothing can be eliminated if the
experiment is performed at low enough temperature, when the
distance between the transverse mode subbands is much larger then
$k_BT$ which for $1T$ hyperfine field is of order of $10 \div
100mK$. The using of high-quality samples will reduce the
influence of scattering. The conditions of homogeneity of the wire
were discussed before.

Let us consider the life time of the NSPI QW. It can be defined by the
following condition: $\left| \varepsilon _{0}^{tr}\left( t_{l}\right)
\right| =k_{B}T^{\ast }$, where $k_{B}$ is the Boltzman constant and $%
T^{\ast }$ is the temperature. Using Eq.(\ref{transverse energy}), we
calculate the time $t_{l}$ for two limiting cases: $T_{1}\ll t_{0}$ and $%
T_{1}\gg t_{0}$. In the first case (the strong relaxation limit) $t_{l}\sim
\frac{T_{1}}{2}\ln \frac{m^{\ast }d^{2}\left( \mu _{B}B_{0}\right) ^{2}}{%
\hbar ^{2}k_{B}T^{\ast }}$. In the second case (the diffusion regime) $%
t_{l}\sim t_{0}\left( \frac{\mu _{B}B_{0}}{k_{B}T^{\ast }}\right) ^{2}$. Let
us estimate the half-width of the wire $\alpha ^{-1}$ at $t\sim t_{l}$. For $%
T^{\ast }=30mK$ and $B_{0}=1T$\ we have $\alpha ^{-1}\sim d$ for $T_{1}\ll
t_{0}$ and $\alpha ^{-1}\sim 20d$ for $T_{1}\gg t_{0}$.

To summarize, a new system for investigation of 1D electron transport - the
nuclear spin polarization induced quantum wire - is proposed. We investigate
the influence of the nuclear spin relaxation and diffusion on the properties
of the electron system. The time dependencies of the electron energy
spectrum and of the conductance of the quantum wire are studied. We expect
that the experimental study of the described system can give some
information, such as the nuclear spin diffusion coefficient $D$ and the
nuclear spin relaxation time $T_{1}$. Furthermore, we note that the method
of local nuclear spin optical polarization allows to create different
low-dimensional nuclear-spin-polarization-induced quantum structures
(quantum dots, rings, wires, etc.) using the same sample and different
illumination masks.

We acknowledge useful discussions with V. Fleurov, V. Ivanov, F.
Peeters and V. Privman. This research was supported in part by the
US National Science Foundation, grants DMR-0121146 and
ECS-0102500, and in part by EU grant IST-2000-29686.

\begin{figure}
 \centering
 \includegraphics[width=18cm]{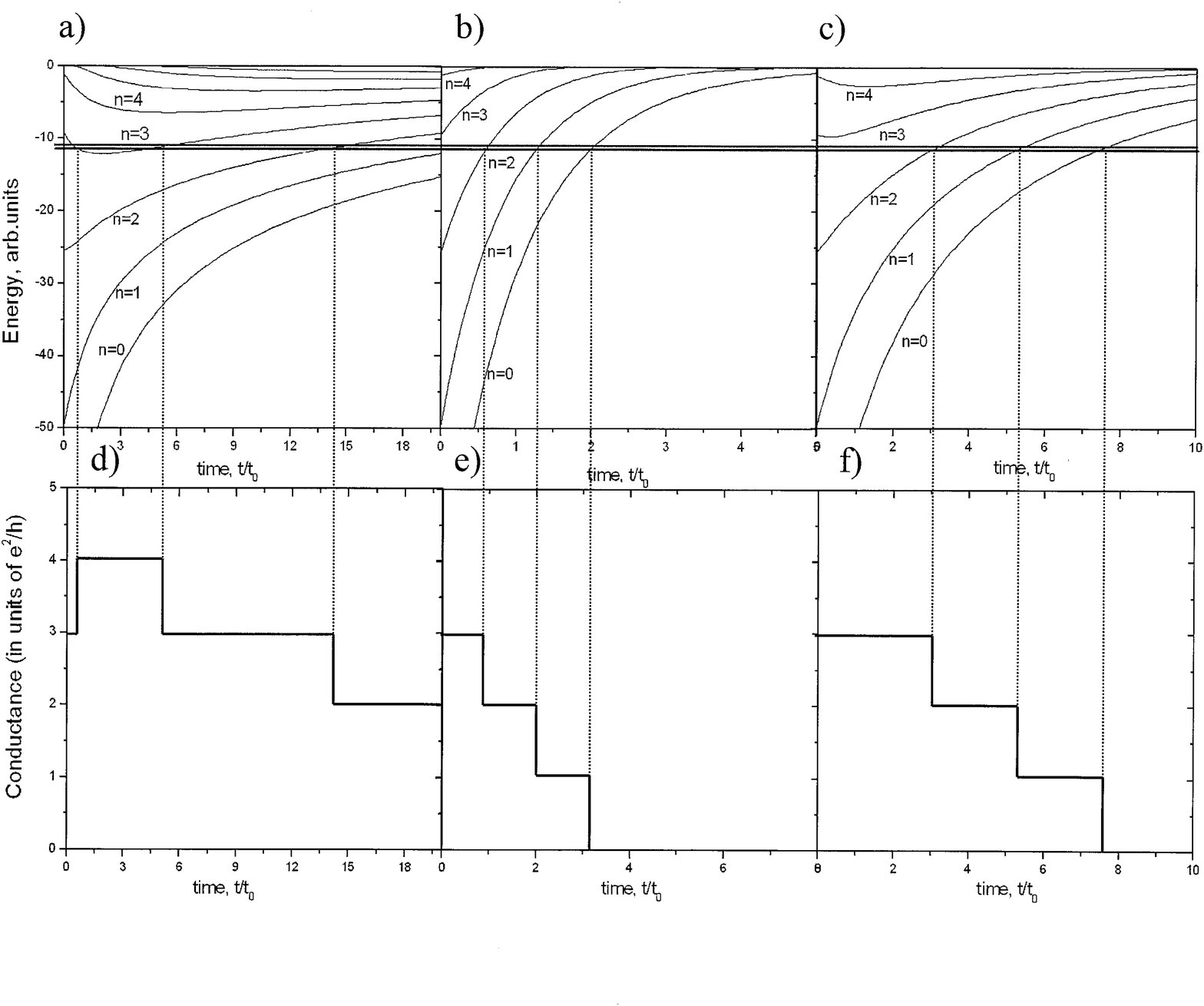}
 \caption{Time evolution of the energy levels in the NSPI QW: a) the
diffusive regime; b) relaxation regime; c) intermediate regime and
d),e),f) respective time dependencies of the conductance. The
parameters of calculations are the same as in Fig.3.}
\end{figure}

\end{document}